\documentclass[twocolumn,prb,superscriptaddress]{revtex4-1}
\usepackage{graphicx} 
\usepackage{bbold}

\begin{document}

\title{A real-space method for highly parallelizable electronic transport calculations}
\author{Baruch Feldman}
\affiliation{Department of Materials and Interfaces, Weizmann Institute of Science, Rehovoth 76100, Israel}
\author{Tamar Seideman}
\affiliation{Department of Chemistry, Northwestern University, 2145 Sheridan Road, Evanston, Illinois 60208-3113, USA}
\author{Oded Hod}
\email{odedhod@tau.ac.il}
\affiliation{Department of Chemical Physics, School of Chemistry, the Raymond and Beverly Sackler Faculty of Exact Sciences, Tel-Aviv University, Tel-Aviv 69978, Israel}
\author{Leeor Kronik}
\email{leeor.kronik@weizmann.ac.il}
\affiliation{Department of Materials and Interfaces, Weizmann Institute of Science, Rehovoth 76100, Israel}

\begin{abstract}
We present a real-space method for first-principles nano-scale electronic transport calculations.  We use the non-equilibrium Green's function method with density functional theory and implement absorbing boundary conditions (ABCs, also known as complex absorbing potentials, or CAPs) to represent the effects of the semi-infinite leads.  In real space, the Kohn-Sham Hamiltonian matrix is highly sparse.  As a result, the transport problem parallelizes naturally and can scale favorably with system size, enabling the computation of conductance in relatively large molecular junction models.  
Our use of ABCs circumvents the demanding task of explicitly calculating the leads' self-energies from surface Green's functions, and is expected to be more accurate than the use of the jellium approximation.  In addition, we take advantage of the sparsity in real space to solve efficiently for the Green's function over the entire energy range relevant to low-bias transport.  
We illustrate the advantages of our method with calculations on several challenging test systems and find good agreement with reference calculation results.
\end{abstract}

\maketitle

\section{ Introduction }

First-principles transport calculations based on the non-equilibrium Green's function (NEGF) approach,\cite{Datta, Landauer, DiVentra} and density functional theory (DFT),\cite{Transiesta, 
Smeagol, KeBaranger, WanT, DiVentraBDT} are popular in the field of molecular electronics and spintronics.  This has fueled a growing demand for large first-principles transport calculations, in order to address increasingly sophisticated nanostructures. 

Many quantum mechanical transport methods make use of localized orbital basis sets so that the system can easily be delineated into electrode and scattering regions.\cite{Transiesta, Smeagol,  KeBaranger, WanT}  
However, for large systems, localized basis set representations may suffer from both scalability and parallelizability problems, especially when diffuse orbitals are used.  Moreover, large basis sets with diffuse, overlapping orbitals have been suggested to lead to ghost transmission \cite{GhostTransRatner} and to related theoretical shortcomings.\cite{ReuterPartition}  
While matching approaches exist for transport in a plane-wave basis,\cite{PlaneWaveTransport} they require a large basis set and are difficult to parallelize.  
Therefore, 
particularly for large-scale transport calculations, it is desirable to consider transport in a real-space basis, which is simple to converge and straightforward to parallelize.  

The advantages of real-space transport calculations are similar to those for electronic structure calculations:\cite{parsec-94, KronikParsec, ParsecBenchmarks, ChelikowskyParsec, Hirose, NatanParsecPBC}  
(\textit{i}) Compared to both plane-wave and localized basis sets, real-space calculations offer a very sparse Hamiltonian.
The method therefore is highly parallelizable, since little communication is required to apply the parallelized Hamiltonian to a trial state vector.  Such a matrix-vector operation is the bottleneck in iterative eigensolution.  This renders the real-space basis useful for computation of large systems.  
(\textit{ii}) Compared to localized or Gaussian basis calculations, the real-space ``basis'' is objective, and convergence with respect to basis size is straightforward. 
(\textit{iii}) Compared to plane-wave calculations, a real-space basis can handle non-periodic systems naturally, and net charges or dipoles present neither conceptual nor calculational difficulties.  
Because 
of its large basis set and unambiguous convergence, a real-space transport method, like the one presented here, may be thought of as a benchmark against which other transport calculations can be tested, as long as any additional approximations (to be described in detail later) are controlled. 
In such a way, a real-space transport method is expected to contribute to the development of transport theory by taking advantage of the growing computational resources.

Previously, Fujimoto, Hirose, Ono, and collaborators \cite{Hirose, FH} have developed real-space formalisms for both electronic structure and transport, and have applied their method to atomic chains \cite{OnoAtomWires} and to some larger test systems.\cite{OnoTransport}  
Extending this theoretical framework, Kong \textit{et al.} \cite{Kong, Kong2007} employed a Lippmann-Schwinger-like matching approach to the real-space transport problem, which allowed them to avoid explicit matrix inversion by instead solving a system of linear equations and taking better advantage of sparsity.

Ono \textit{et al.} and Kong \textit{et al.} expressed the Green's function $G$ in real space, leading to very large computational demand.  Both groups, in practice, restricted their computations, for systems larger than atomic chains, to the jellium model for the leads (although self-energy calculations are possible in their framework), and to a minimal atomic description of the extended molecule.\cite{OnoTransport, Kong, Kong2007}
Therefore, there is room for refinement of real-space transport methods in order to make them efficient for fully atomistic calculations on highly extended systems.  
Naturally, to make the method competitive, it is desirable to make the best possible use of the sparsity of the Hamiltonian and other matrices in this basis.  

In this paper, we present a real-space, highly parallel method for first-principles Landauer electronic transport calculations, using absorbing boundary conditions \cite{Neuhauser-ABCs, RissMeyer-ABCs, SeidemanMiller, Henderson-ABCs, Hod-ABCs} (ABCs, also known as complex absorbing potentials, or CAPs) to mimic the proper outgoing-wave boundary conditions.  We base our implementation, which we call TRANSEC, on the PARSEC (Pseudopotential Algorithm for Real-Space Electronic Calculations) real-space 
DFT code.\cite{parsec-94, KronikParsec}  
We note that while some approaches to electronic transport calculations do not rely on pseudopotentials,\cite{GFLEUR} 
 their use has been well-justified at the level of electronic structure calculations.\cite{ParsecBook, Pickett, Chelikowsky}  Because electronic transport is governed by the underlying electronic structure, the use of pseudopotentials in electronic transport calculations is widely accepted as reliable.\cite{Transiesta, Smeagol, WanT}

Instead of directly solving for a large sub-matrix of the Green's function $G$, we have chosen to take advantage of real- and energy-space sparsity by iteratively diagonalizing the sub-space of the Kohn-Sham (KS) Hamiltonian most relevant for transport.  This approach accurately yields the relevant sub-matrices of $G(E)$ for a dense set of energies $E$ in significantly less time than full solution at a single energy.  
In addition, our use of ABCs allows the treatment of large, realistic contact regions from first principles.  

The next section presents our method, and in Sec.~\ref{sec:Results} we present test calculations with our method on large Au(111) electrodes with an Au atomic contact and the benzene dithiol molecule, which are found to be in good agreement with analytical expectations.  
We conclude in Sec.~\ref{sec:Conclude} with a discussion of the method's strengths and weaknesses, as well as planned future improvements.

\section{Computational methods \label{sec:Method}}

\subsection{Landauer approach}

We use the Landauer approach \cite{Datta, Landauer} to calculate the current flowing through a molecular junction, $I$, 
\begin{equation}
I = \frac{2e}{h} \int_{-\infty}^\infty T\left( E \right) \: \left[ f\left(E-\mu_L\right) - f\left(E-\mu_R\right) \right] \: dE \: ,
\label{eq:Landauer}
\end{equation}
where $e$ is the electronic charge, $h$ is Planck's constant, $T(E)$ is the transmission probability through the junction at energy $E$, $f$ is the Fermi-Dirac distribution function and $\mu_{L,R} = \pm eV/2$ are the chemical potentials of the left and right electrodes at bias $V$, respectively.  

The transmission function $T$ is given by \cite{SeidemanMiller,Datta,DiVentra} 
\begin{equation}
T\left( E \right) = \mbox{Tr}\left\{ G^r\left( E \right) \: \Gamma_R \: G^a\left( E \right) \: \Gamma_L \right\} ,
\label{eq:T}
\end{equation}
where 
$\Gamma_{L, R}$ are matrices that couple electrons to the left and right electrodes,
$G^r$ is the retarded single-particle Green's function, 
\begin{equation}
G^r\left( E \right) \equiv \left[ \: E \mathbb{1} - H_{op} - i \eta \: \right]^{-1},
\label{eq:G}
\end{equation}
$G^a = G^{r \dagger}$ is the advanced Green's function, $H_{op}$ is the matrix representation of the Hamiltonian for the {\em open} system which, in our case, is represented by the KS-DFT single particle picture, and $\eta \rightarrow 0^+$.  
Eq.~(\ref{eq:G}) implies that the transport problem requires a large (in principle infinite) matrix inversion which may become a bottleneck in real-space calculations.  

We next discuss our approach to treating $H_{op}$ based on a DFT framework.  

\subsection{Absorbing boundary conditions (ABCs) \label{sec:ABCs}}

Typically, first-principles electronic transport calculations in the NEGF formalism make use of self-energies computed from surface Green's functions \cite{Datta, Transiesta, KeBaranger} to incorporate the effects of the semi-infinite leads in $H_{op}$.  However, computing self-energies can be time-consuming, and must be done independently at each energy $E$.  
To reduce this burden, several previous works, and in particular the real-space transport studies of Ono \textit{et al.} and Kong \textit{et al.},\cite{OnoTransport, Kong, Kong2007} used the jellium approximation to represent semi-infinite electrodes (although, in the latter case, their formalism can also be used with self-energies).  

We wish to use realistic electrode models, while also avoiding the computational cost of explicitly calculating self-energies.  
To this end, we adopt the absorbing boundary condition (ABCs) method, where local ABCs are added to the KS effective potential, $V_{KS}$, near the edges of the lead models.  
ABCs have been previously used in combination with a variety of approaches for electron transport and for effectively mimicking self-energies in non-equilibrium Green's function approaches.\cite{Neuhauser-ABCs, RissMeyer-ABCs, SeidemanMiller, Henderson-ABCs, Hod-ABCs, ABC-SelfEn, Roi-ACDC, ABC-Transient, ABC-multiterm, ABC-review}
In this approach, $H_{op}$ is represented by  
\begin{equation}
H_{op} \equiv H_{KS} - i \Gamma,
\label{eq:Hop}
\end{equation}
where $H_{KS}$ is the usual Kohn-Sham Hamiltonian for the (finite) model system and $\Gamma=\Gamma_L + \Gamma_R$.  For simplicity we choose a Gaussian form for the ABCs, 
\begin{equation}
\Gamma_{L, R}(x,y,z) = 
\Gamma_0 \: e^{- {(z-z_{L, R})^2} / \sigma_{L, R}^2 }  \; , 
\label{eq:ABC}
\end{equation}
where $x,y,z$ are the quantum mechanical position \emph{operators}, and therefore $\Gamma$ is diagonal in the real-space representation.  The ABC strength $\Gamma_0$ and characteristic width $\sigma_{L,R}$ are parameters that need to be tuned to absorb incoming Bloch waves in a given electrode structure and energy range.  The location $z_{L,R}$ of the ABC center is typically set to the edge of the electrode.  
Here we choose, without loss of generality, the main axis of the leads to be aligned along the $z$ direction.  
Because $H_{KS}$ is Hermitian and purely real 
and because $\Gamma_{L,R}$ are purely diagonal in real space, $H_{op}$ is a {\em complex symmetric} (non-Hermitian) operator: 
\begin{equation}
H_{op}^\dagger = H_{op}^* \: . 
\label{eq:complex-sym}
\end{equation}

If the transmission function $T\left(E\right)$ only needs to be investigated over a small energy range (low bias), the ``wide-band limit approximation,'' \cite{GhostTransRatner} namely, taking the ABCs as independent of energy, applies.  (This is quantified, for example, by numerical results in Ref.~\cite{RissMeyer-ABCs}.)  Moreover, because the ABCs are diagonal and purely imaginary, they can play the role of the anti-Hermitian part of the self-energies, and therefore are used as the coupling matrices $\Gamma_{L, R}$ of Eq.~(\ref{eq:T}).\cite{KeBaranger,Henderson-ABCs}  

Aside from the advantages relative to jellium or self-energy treatments of the open system emphasized above, the use of ABCs results in an advantage in the implementation and scaling of real-space $T\left(E\right)$ calculations.  
Applying the complex symmetry of 
$G^r$, $G^a$ and the fact that $\Gamma_{L,R}$ are diagonal to
the formula for transmission, Eq.~(\ref{eq:T}), 
yields (see Appendix): 
\begin{equation}
T(E) = \sum_{i \in L, j \in R} |G^r_{i j}(E)|^2 \: \Gamma_{R j j} \: \Gamma_{L i i} \;, 
\label{eq:T-realspace}
\end{equation}
where $i, j$ are real-space grid points in the left and right ABCs, respectively.\cite{SeidemanMiller}  
Therefore, only a sub-matrix of $G^r$ coupling $\Gamma_L$ to $\Gamma_R$ needs to be computed.  This results in advantageous scaling of the $T\left(E\right)$ calculation with system length when the length 
is increased beyond the extent $\sim\!\sigma_{L, R}$ of the ABCs.

\subsection{Complex energy diagonalization \label{sec:cplx-diag}}

At this point, the remaining task is to obtain (the relevant subblock of) the Green's function via Eq.~(\ref{eq:G}). This implies the inversion of a matrix that is closely related to the KS Hamiltonian matrix. In real space, the KS Hamiltonian matrix is highly sparse, with the only off-diagonal elements given by the high-order finite-difference expansion for the Laplacian operator and the non-local part of the pseudopotentials (which are nevertheless constrained by a cutoff radius).\cite{parsec-94}  This sparseness leads to good parallelizability in the diagonalization of the KS matrix, because communication is only required when off-diagonal blocks connect grid elements on different processors.\cite{KronikParsec,ParsecBenchmarks,ChelikowskyParsec,NatanParsecPBC}   Unfortunately, these advantages do not carry over to direct inversion because the inverse of a sparse matrix is generally non-sparse.

To take advantage of both the sparseness of $H_{KS}$ and the small energy window around $E_F$ that governs low-bias transport, 
we avoid direct inversion by partially diagonalizing $\left( G^r \right)^{-1} = (E \mathbb{1} - H_{KS} + i \Gamma)$ using an iterative eigensolver.  This approach allows us to find eigenpairs near $E_F$ with a computational cost that can scale very slowly with the size of $H_{KS}$.  Because $\left( G^r \right)^{-1}$ is non-Hermitian, we should note that diagonalizability is not strictly guaranteed, but is expected in numerical practice.\cite{SantraCederbaum}  

First, we define $U$ as the matrix whose columns $U_i$ are right eigenvectors of $H_{op}$ (and therefore of $\left( G^r \right)^{-1}$):
\begin{equation}
\left( H_{KS} - i \Gamma \right) \: U_i \; = \; \epsilon_i \: U_i \: , \; \epsilon_i \: \in \: \mathbb{C} \:.
\label{eq:complex-energy-evp}
\end{equation}
Complex conjugation of (\ref{eq:complex-energy-evp}) requires that $U_i^*$ are the eigenvectors of $H_{op}^\dagger$ (and therefore of $[G^a]^{-1}$) with eigenvalues $\epsilon_i^*$.  Note that due to the complex symmetry (\ref{eq:complex-sym}) of $H_{op}$,  
$U$ is \textit{complex orthogonal}:\cite{SantraCederbaum}
\[
U^\dagger = (U^{-1})^* \: .
\]  
As a result, the basis $\{ U_i \}$ is bi-orthogonal, rather than orthonormal under the standard positive-definite inner product.\cite{SantraCederbaum,CSYM}  

The 
representation of $G^r$ in this basis is diagonal: 
\begin{eqnarray}
\nonumber 
\tilde{G^r}(E) \; \equiv \; U^{-1} \: G^r(E) \: U = \: \left[ U^{-1} \: \left(E \mathbb{1} - H_{op}\right) \: U \right]^{-1} \: 
\\ 
= \: \mathrm{diag} \{ \: 1/\left( E - \epsilon_i \right) \: \} \: , \;
\label{eq:G-diag}
\end{eqnarray}
where the notation $\mathrm{diag}$ refers to a diagonal matrix with the given elements.  
Having found $U$ and $\epsilon$, our task 
is reduced to computing 
\begin{eqnarray}
\nonumber G^r(E) \: = \: U \: \tilde{G^r}(E) \: U^{-1} \; , \\
G^a(E) \: = \: (G^r(E))^* \; ,
\label{eq:G-from-diag}
\end{eqnarray}
with $\tilde{G^r}$ the \textit{diagonal} matrix given in Eq.~(\ref{eq:G-diag}).  
Therefore, once $U$ and $\epsilon$ are calculated, we are equipped to find $G$ over a \textit{whole range}
of energies $E$ with little additional computation.  

Let us consider how this facilitates the computation of the transmission function $T(E)$.  
In the bases $U$, $U^*$, subsituting Eq.~(\ref{eq:G-from-diag}) into Eq.~(\ref{eq:T}) and applying the cyclical property of the trace gives
\begin{equation}
T\left( E \right) = \mbox{Tr}\left\{ \tilde{G^r}\left( E \right) \: \tilde{\Gamma}_R \: \tilde{G}^a\left( E \right) \: \tilde{\Gamma}_L \right\} ,
\label{eq:T-tilde}
\end{equation}
where we have made the new definitions: 
\begin{eqnarray}
\nonumber \tilde{\Gamma}_{L} \equiv \: U^{\dagger} \: \Gamma_{L} \: U \: , \\ 
\tilde{\Gamma}_{R} \equiv \: U^{T} \: \Gamma_{R} \: U^* \; .  
\label{eq:Gamma-tilde}
\end{eqnarray}
While $G$ has become diagonal, $\Gamma$ no longer is.  
However, only the blocks of $G$ and $\Gamma_{L,R}$ that correspond to the same eigenpairs multiply each other, as can be seen from explicitly evaluating Eq.~(\ref{eq:T-tilde}), which gives the equivalent of Eq.~(\ref{eq:T-realspace}) in the new basis (see Appendix):
\begin{equation}
T(E) = \sum_{i,j}^N  \frac{\tilde{\Gamma}_{L i j} \: \tilde{\Gamma}_{R j i}}{(E-\epsilon_j)(E-\epsilon^*_i)} \;, 
\label{eq:T-tilde-realsp}
\end{equation}
with $N$ the dimension of the real-space grid.  
The denominator of Eq.~(\ref{eq:T-tilde-realsp}) implies that eigenvalues $\epsilon_i$ far from $E$ contribute only weakly to $\tilde{G^r}$ and $\tilde{G}^a$.   Thus, we need only the blocks of 
$\tilde{\Gamma}_{L,R}$ with indices corresponding to eigenvalues
near the bias window
\[
E_F - \frac{e V}{2} \; < \; E \; < \; E_F + \: \frac{e V}{2} \: . \]  
Evaluating (\ref{eq:Gamma-tilde}) explicitly and using the fact that $\Gamma_{L,R}$ are diagonal in real space, yields
\[
\tilde{\Gamma}_{L i j} = \sum_k^N \: U^*_{k i} \: U_{k j} \: \Gamma_{L k k} \: , \; 
\tilde{\Gamma}_{R j i} = \sum_k^N \: U^*_{k i} \: U_{k j} \: \Gamma_{R k k} \: .   
\]
Therefore, in order to find the needed blocks of $\tilde{\Gamma}_{L,R}$, we also need to compute \textit{only} the restricted set of eigenvectors $U_i$ 
that correspond to the eigenvalues $\epsilon_i$ near the bias window.  

Thanks to this fact, we need to diagonalize iteratively just a small fraction $p$ (typically $\sim$1\%) of the total space.   
Strictly, $p$ should be tested for convergence of $T(E)$, as we have observed that missing eigenpairs tend to produce a pole-like behavior in Eq.~(\ref{eq:T-tilde-realsp}).  [This is understood because the LHS of Eq.~(\ref{eq:T-tilde-realsp}) is typically a smooth function, while the RHS is a sum of terms with poles.]  
We performed this test for the calculations presented in Sec.~\ref{sec:Results}.  
But as a rule of thumb, since $U$ is complex orthogonal, we may expect that typical elements in 
$\tilde{\Gamma}_{L,R}$ and $\Gamma_{L,R}$ are of comparable magnitude,
and one may simply exclude eigenpairs for which 
\[
|E - \epsilon_j|^2 \; \gg \; \Gamma_0^2 \: ,
\]
with $\Gamma_0$ the ABC strength as in Eq.~(\ref{eq:ABC}).  
Even this approximate criterion cannot be predicted with certainty because the complex eigenvalues $\epsilon$ are unknown in advance, but it can be checked after the diagonalization.  Alternatively the (real) KS eigenvalues can be inserted in this test as approximations of the complex ones.  This is valid because, through much of the simulation cell, $i \Gamma$ is a small perturbation to the KS potential 
$V_{KS}$, so the complex eigenvalues $\epsilon$ correspond roughly to the (real) KS eigenvalues, $|\epsilon| \sim |\epsilon_{KS}|$, and generally $|\Im\{\epsilon_i\} | \: \ll \: |\epsilon_i|$.

\subsection{Computational details} 

In our real-space transport code, TRANSEC, we implemented the complex eigensolution of Eq.~(\ref{eq:complex-energy-evp}) using the complex iterative diagonalization routines in the package ARPACK.\cite{Arpack, CSYM-scaling}  

In the computations presented in Sec.~\ref{sec:Results}, we used typically a grid spacing of 0.6 to 0.7 $a_0$  [where we use atomic units, $a_0$ = 1 bohr].  We tested these values for convergence of the KS eigenvalues, and, in the case of the atomic chains, of $T(E)$, as well.  Note that the scaling of the Hamiltonian dimension $N$ with grid spacing $h$ is $N \propto h^{-3}$, but the time for fully solving for $G$ scales even more strongly, thus making use of sparsity crucial.  
To this end, we used a fraction of the total number of complex eigenpairs, $p \sim$ 1\% to 2.5\%, and also tested these for convergence of $T$.  

The computational cost of the iterative diagonalization of Eq.~(\ref{eq:complex-energy-evp}) scales like $N n_r^2$, where $N$ is the dimension of the Hamiltonian, and $n_r = p N$ equals the total number of eigenpairs found.\cite{PRIMME}  Here only a single factor of $N$ comes from applying the Hamiltonian, which is a critical source of parallelization in real-space methods.\cite{parsec-94, KronikParsec}  Note that $N$ depends on both the system volume (number of atoms) 
and the grid spacing, whereas $n_r$ typically scales only with the number of electrons in the system.  

We used the local density approximation \cite{CA} for the atomic chain tests described below, involving C and H, and the generalized gradient approximation of Perdew, Burke, and Ernzerhof (PBE) \cite{PBE} for the chain calculations involving Au, and for the larger calculations with Au(111) nanowire electrodes.  

\section{Results and Discussion \label{sec:Results}}

\subsection{Atomic chain tests \label{sec:atom-chains}}

\begin{figure}
\includegraphics[scale=1]{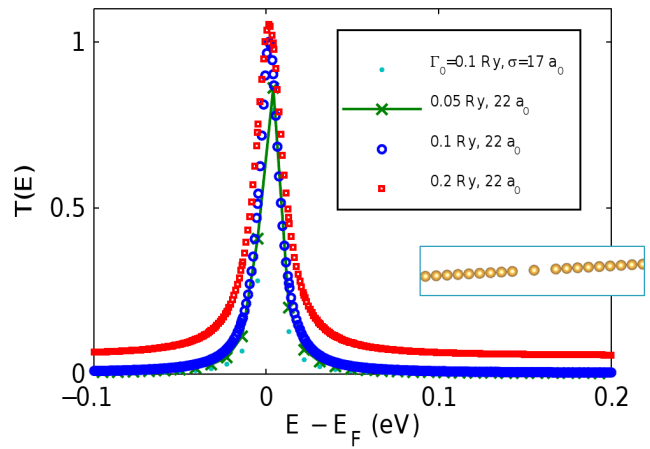}
\caption{Robustness of $T(E)$ with respect to ABC (equivalently, CAP) parameters for a monatomic Au chain/atom/chain structure (shown as inset).   
First parameter in the legend is $\Gamma_0$, the ABC strength, second parameter is $\sigma$, the ABC width along the transport direction.  Lines are shown to guide the eye only.  
Note that the behavior seen in the last dataset (red squares) may be caused by different convergence with respect to the fraction $p$ of calculated eigenpairs than for the other ABC parameter sets shown.  
\label{fig:ABC-insensitivity} }
\end{figure}

We start by considering simple atomic chain models to test the performance of our method.  First, we calibrate the ABC parameters (height $\Gamma_0$ and width $\sigma_{L,R}$; peak locations $z_{L,R}$ are set to the ends of the chains) by comparing the transmission probability to an analytical expression and by identifying regions where $T(E)$ is robust against small modifications in the parameters.
Typical results are shown in Fig.~\ref{fig:ABC-insensitivity} for an Au monatomic chain structure with interatomic spacing of 5.5 $a_0$ and a single Au atom ``device'' separated from the electrodes by a gap of twice the atomic spacing.  
With an appropriate choice of ABC parameters it is found that the transmission probability of this system is robust against changes of 100\% in the ABC height and $\sim$23\% in its width.
This insensitivity 
simplifies calibrating the ABCs for a given set of electrodes, as well as making the ABC tuning more predictive.  Once calibrated for a single well-known lead model, the ABC parameters can, in principle, be used with any (extended) molecule inserted between the calibrated electrodes.

\begin{figure}
\includegraphics[scale=0.8]{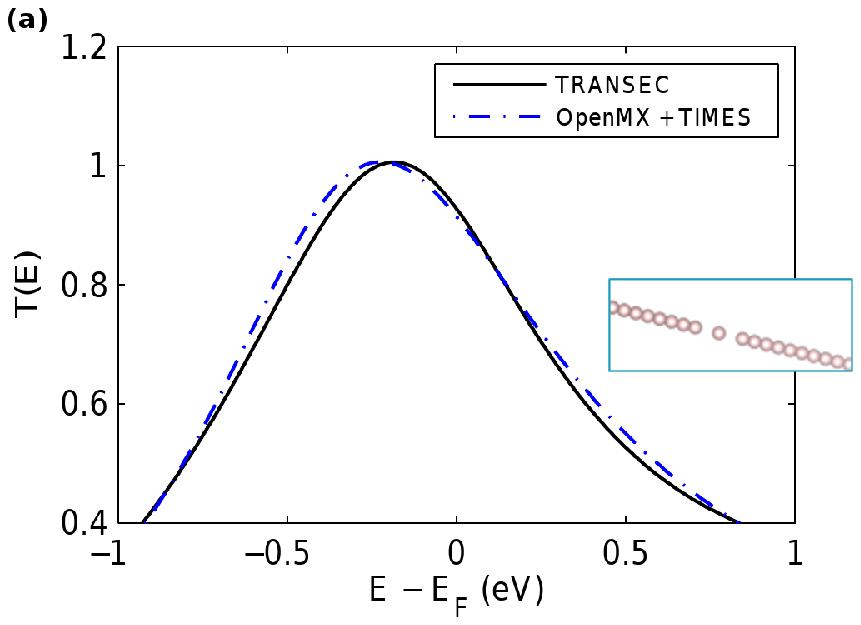} \\
\includegraphics[scale=0.8]{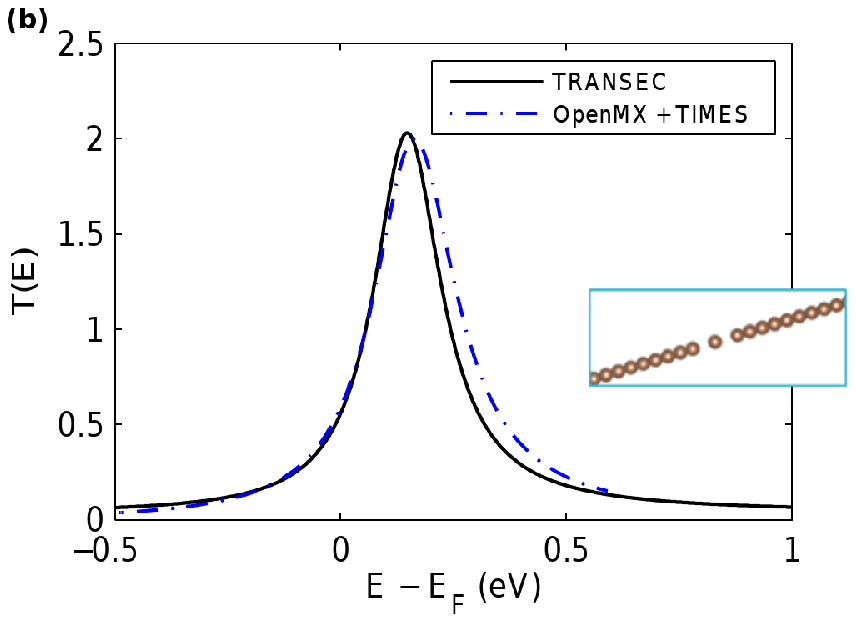} \\ 
\includegraphics[scale=0.8]{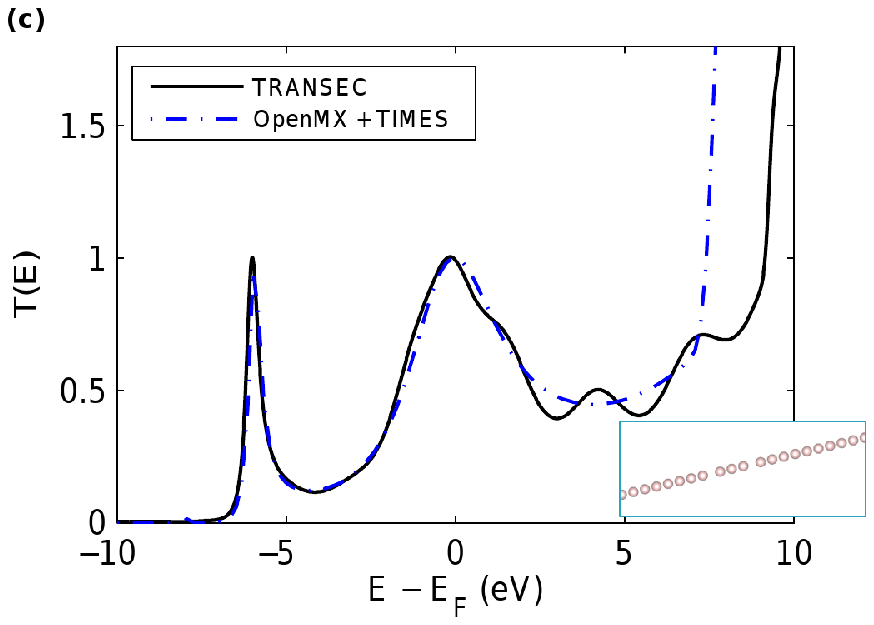} 
\caption{TRANSEC calculated transmission probability curves in the vicinity of the Fermi energy for monatomic chain/device/chain configurations with device = (\textit{a}) H, (\textit{b}) C, and (\textit{c}) 3H (structures shown in insets), together with reference results obtained using OpenMX + TIMES \cite{OMX, TIMES}.  
\label{fig:atomic-chain-results}
}
\end{figure}

Figure \ref{fig:atomic-chain-results} shows results for hydrogen and carbon monatomic chains with a single- or several-atom ``device'' as the scattering region, separated by a gap (larger than the interatomic spacing in the leads) from the atomic chain leads.  We have chosen these systems because they are readily computable with both TRANSEC and localized-orbital reference transport codes, and because their physics is well understood.  
For hydrogen, the inter-atomic spacing used was 2 $a_0$ and the electrodes were 12 atoms long (note that both hydrogen systems have identical electrodes, and therefore use the same ABC parameters).  For carbon, the spacing was 2.6 $a_0$ and the electrodes were 14 atoms long.  The gaps were (\textit{a}) 4 $a_0$ (\textit{b}) 4.7 $a_0$, and (\textit{c}) 3 $a_0$, giving a total chain length of (\textit{a}) 52 $a_0$, (\textit{b}) 77 $a_0$, and (\textit{c}) 54 $a_0$.   
The ABC parameters used were $\Gamma_0=$ 265 mRy and $\sigma=$ 6 $a_0$ for both hydrogen chain systems, and $\Gamma_0=$ 265 mRy, $\sigma=$ 10.4 $a_0$ for the carbon chain.  
All calculations used a grid spacing of 0.6 $a_0$, 
giving Hamiltonian dimensions $N$ from 13,000 to 22,000.  
Also shown are reference calculations 
using the TIMES transport code \cite{TIMES} to compute $T(E)$ in the linear response regime based on OpenMX \cite{OMX} electronic structure.
The OpenMX calculation used a basis set of 17 orbitals per atom for both the hydrogen and carbon chains.

As is well-known, the transmission probability for a single energy level coupled weakly to two baths can be modeled analytically by a Lorentzian, where the peak width depends on the coupling strength.\cite{Datta}  
Because the device is coupled weakly to the electrodes, we expect a $T(E)$ peak near $E=E_F$ of height equal to the number of conductance channels (1 for H, 2 for C) and width dependent on the electrode-device distance.    
As can be seen from the figure, the calculations agree well with these expectations and with the OpenMX + TIMES results.  

We note that in some test cases, agreement is found to be worse 
than shown in Fig.~\ref{fig:atomic-chain-results} 
for energies away from $E_F$.  We found that this is caused by basis set convergence in the DFT part of the atomic orbital calculation, leading to disagreement in the band structures.  
We found that 
improving the convergence of the OpenMX + TIMES calculations typically improves its agreement with the real-space results.

\subsection{Au(111) nanowire/Au atom/Au(111) nanowire \label{sec:Au-nanowire}}

Following the calculations for simple atomic chain models, we now turn to present the results of our benchmark calculations for a larger Au(111) nanowire electrode/atom/electrode junction.  
We chose this system to benchmark our method's capabilities because its large size and the presence of Au atoms make it challenging, particularly when using localized orbital representations.  Because it has a near-continuum of energies in the electrodes and an isolated atom with discrete levels as the ``device,'' $T(E)$ is expected to display a Lorentzian-like peak near $E_F$ with width dependent on the electrode-atom gap, as for the test cases studied in Sec.~\ref{sec:atom-chains}, and can serve as a benchmark.  

\begin{figure}
\includegraphics[scale=0.3, angle=-90]{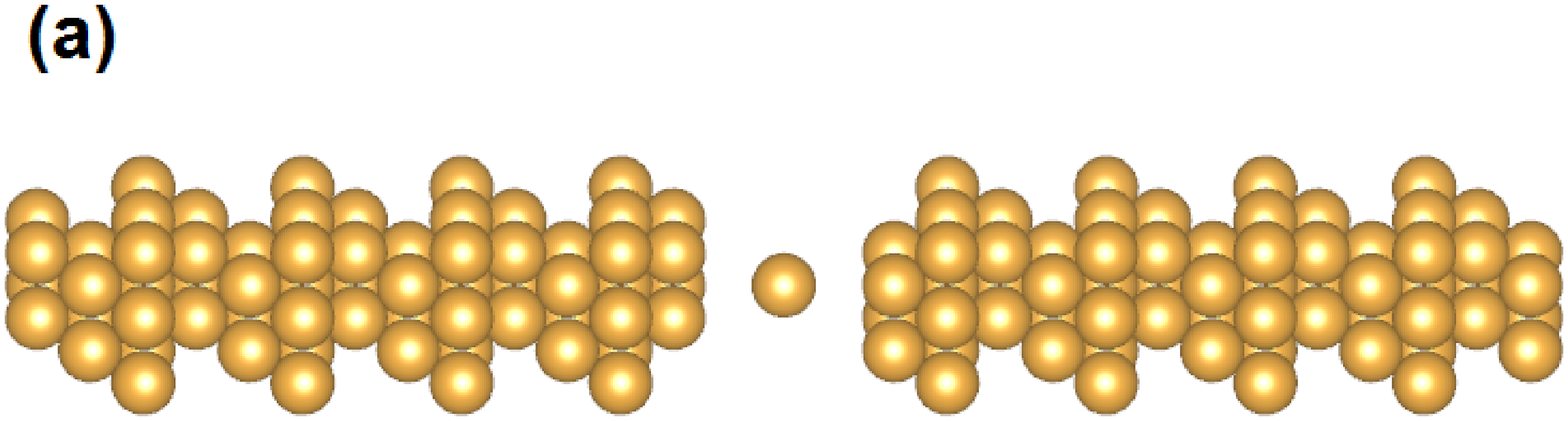} \\
\includegraphics[scale=0.3, angle=-90]{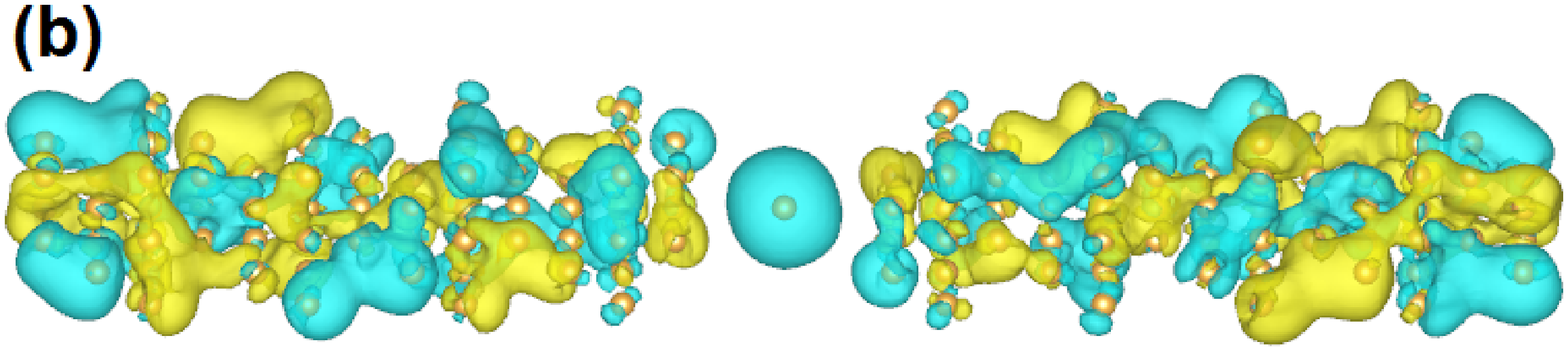}
\caption{(a) Geometry of the Au(111) nanowire/atom/nanowire system.  The Au atom is weakly coupled to two Au(111) nanowire leads (separated by 9 $a_0$ on each side).    
(b) Isosurface of a low-lying partially-filled orbital of the system shown in (a), with isovalue $3\cdot10^{-3} \: a_0^{-3/2}$.   \label{fig:Au111} }
\end{figure}

The test geometry used is shown in Fig.~\ref{fig:Au111}.  
The electrodes were formed of Au(111) nanowires with 72 atoms each.  
The simulation cell's lateral dimensions were 25 $a_0$~$\times$ 20 $a_0$, and the dimension along the transport axis was 161 $a_0$.  
We used a norm-conserving Troullier-Martins pseudopotential \cite{TroullierMartins} for Au with electronic configuration of $5d^{10}6s^16p^0$ and \textit{s}/\textit{p}/\textit{d} cutoff radii of 2.77/2.60/2.84 $a_0$.\cite{DoronLeeor} 
The grid spacing was $h=$ 0.7 $a_0$, giving a total Hamiltonian dimension $N \approx$ 234,500.  
Using a fraction $p=$ 1.25\% of the total eigenpairs, the TRANSEC calculation took approximately 41 hours of wall time running on 24 cores (980 core-hours) of Intel E5-2630 at 2.3 GHz clock speed with a total of 128 GB RAM. 

We used Gaussian ABCs (eq.~(\ref{eq:ABC})) centered on the outermost Au monolayers, with $\Gamma_0=$ 100 mRy and $\sigma =$ 8.5 $a_0$ (so that the ABC has sufficiently decayed before the central atom).  
To calibrate the ABCs, and as a test of our method applied to a large system, we placed a single Au atom in the device region, separated from the electrodes by 9 $a_0$~on each side.  Because the electrodes are large enough to have a virtually continuous KS eigenvalue spectrum, and since the central atom is weakly coupled to the leads, transmission is limited by the energy levels of the (isolated) device atom.  Therefore, we expect $T(E)$ to display a narrow peak with height 1 near $E=\epsilon_{KS}$, where $\epsilon_{KS}$ is the (real) KS eigenvalue of an orbital isolated on the device atom, and with width dependent on the electrode-atom gap.  

Note that even in the limit of zero physical coupling between the device atom and the electrodes, KS-DFT requires all sub-systems to be filled to a common Fermi level, $E_F$.  Therefore, $E_F$ of the full nanowire/atom/nanowire system, as computed by KS-DFT, must lie between the HOMO and LUMO KS eigenvalues for the (isolated) device atom.  So an orbital largely localized on the device atom is expected to be found within $\sim1$ eV of $E_F$.  
One of the first few partially-occupied orbitals of the electrode/atom/electrode system is such an orbital with significant amplitude on the isolated atom, as shown in Figure \ref{fig:Au111}(b).  Transmission through this orbital is expected to be responsible for a Lorentzian peak in $T(E)$ near $E = E_F$, like in the case of the atomic chains.  

\begin{figure}
\includegraphics[scale=0.32, angle=-90]{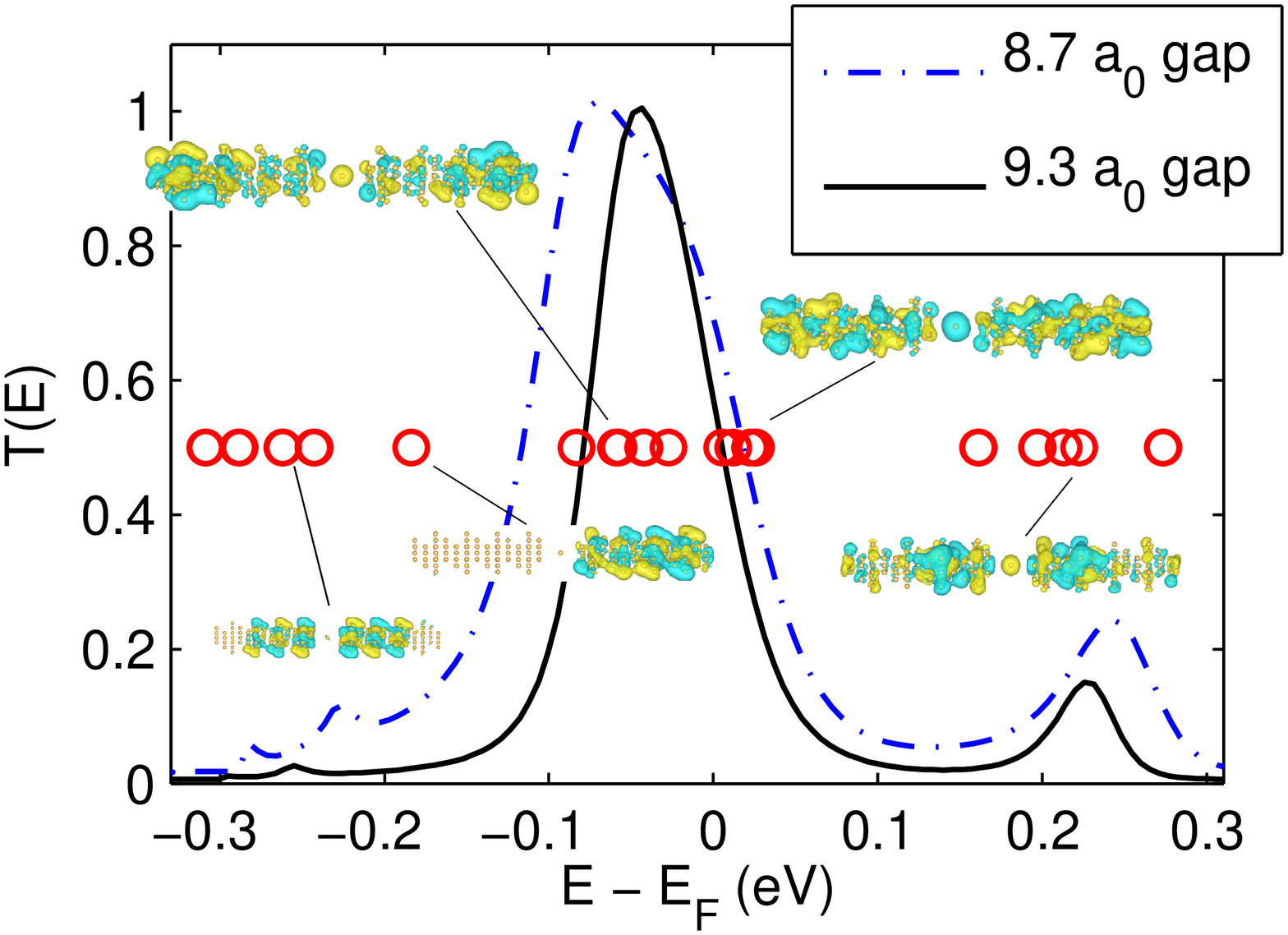}
\caption{(Solid curve) TRANSEC calculated results for transmission, $T$, for the Au(111) nanowire/atom/nanowire system shown in Fig.~\ref{fig:Au111}, together with the locations of the (real) KS eigenvalues (red circles), and isosurfaces of representative KS orbitals.  
\label{fig:Au-results} 
Also shown (dashed curve) is the effect on $T(E)$ of  
a larger gap between nanowire and central Au atom.  A larger gap corresponds to weaker lead-atom coupling, and displays a narrower main peak. } 
\end{figure}

Figure \ref{fig:Au-results} shows the calculated $T\left(E\right)$ results.
The calculation is in good agreement with our prediction: as expected, a peak appears near $E_F$ with a height of 1.  We also verified, by varying the electrode-atom gap distance, that the peak width displays the expected dependence on the electrode-device coupling, as shown in the figure.  To explain further the features seen in $T(E)$, we also plot in Figure \ref{fig:Au-results} the location of the (real) KS eigenvalues together with several representative orbitals.  Typically, large $T(E)$ peaks coincide with delocalized KS molecular orbitals that bridge the Au atom ``device'' to the leads.  More localized orbitals contribute much less to the transmission.  
Note, for example, that the KS orbital near $E_F - 0.18$ eV vanishes in the left electrode and therefore contributes negligibly to $T(E)$.

\subsection{Benzene dithiol molecule \label{sec:BDT}}

\begin{figure}
\includegraphics[scale=0.41, angle=-90]{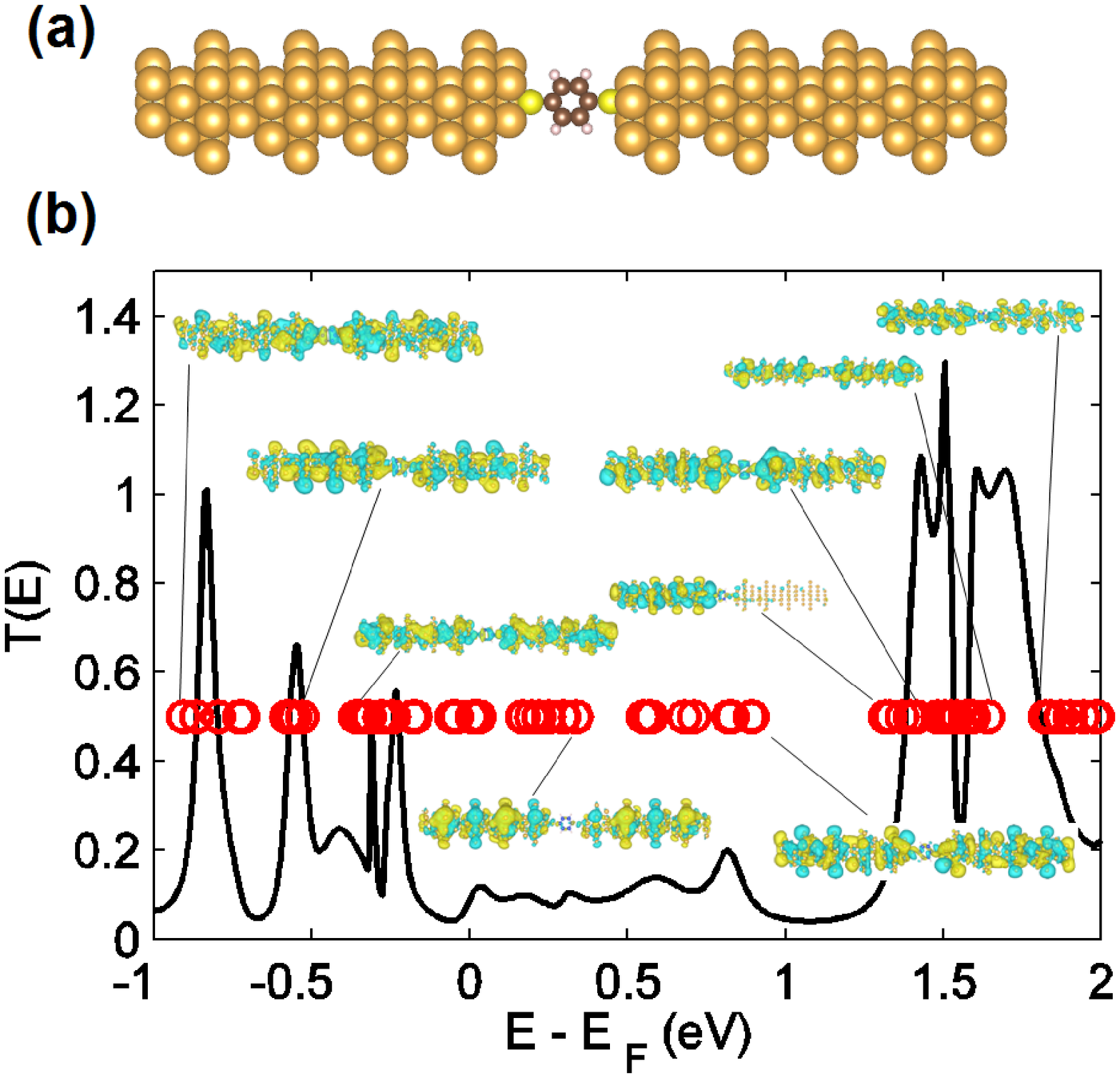}
\caption{(a) Geometry of the Au(111) nanowire/BDT/nanowire system.  The plane of the BDT molecule is oriented along the nanowire axis, and the separation between S and the Au(111) plane is 3.2 $a_0$, as in the study of Stokbro \textit{et al.}.\cite{Stokbro-BDT}  
(b) TRANSEC calculated results for transmission, $T(E)$, for the Au(111) electrode/benzene dithiol molecule/electrode system, together with the locations of the (real) KS eigenvalues (red circles), and isosurfaces of representative KS orbitals.  \label{fig:BDT} }
\end{figure}

Having validated our approach for atomic scattering regions, we now turn to demonstrate it on molecular scattering regions.  
We have applied the method to compute $T(E)$ for the benzene dithiol (BDT) molecule between Au electrodes, shown in Fig.~\ref{fig:BDT}, a system that has been extensively studied computationally.\cite{DiVentraBDT, Stokbro-BDT, Emberly, Li-BDT, Bauschlicher, Strange}  The results of the $T(E)$ computation are shown in Fig.~\ref{fig:BDT}, together with the (real) KS eigenvalues and representative orbital isosurface plots.  The Au(111) nanowire electrodes are the same as those we used in the Au nanowire/atom/nanowire configuration shown previously.  We note that we did not have to re-calibrate the ABC parameters from that calculation, so these results (as well as the H chains in Figs.~\ref{fig:atomic-chain-results}(a) and (c)) provide an illustration of the transferability of ABC parameters for different (extended) molecules using the same lead models.  
We used norm-conserving Troullier-Martins pseudopotentials with 
\textit{s}/\textit{p}/\textit{d} cutoff radii of 1.69/1.69/1.69 $a_0$ for S, 
\textit{s}/\textit{p} cutoff radii of 1.46/1.46 $a_0$ for C, and \textit{s} cutoff radius of 1.28 $a_0$ for H.  

We based our geometry (Fig.~\ref{fig:BDT}) on the structure considered by Stokbro \textit{et al.},\cite{Stokbro-BDT} including the Au-S separation of 3.2 $a_0$ along the transport direction, and the placement and angle of the BDT molecule relative to the Au FCC (111) face.  However, there remain important differences, such as Stokbro \textit{et al.}'s use of periodic boundary conditions in the lateral dimension, thus modeling the transport through a molecular monolayer while we address single molecule transport.  Qualitatively, our $T(E)$ results show peaks around $E_F - 1$ eV and several eV above $E_F$, as do theirs, and the remaining quantitative differences are within the spread of results reported in other computational studies.\cite{Stokbro-BDT, Emberly, Li-BDT, Bauschlicher, Strange}  
 To validate further our $T(E)$ curve for the given DFT-computed electronic structure, we performed the (real) KS eigenvalue analysis shown in Figure~\ref{fig:BDT}, which shows agreement between the locations of the peaks and the KS eigenvalues corresponding to delocalized molecular orbitals.

\subsection{C$_{60}$ molecule \label{sec:C60}}

\begin{figure}
\includegraphics[scale=0.43,angle=-90]{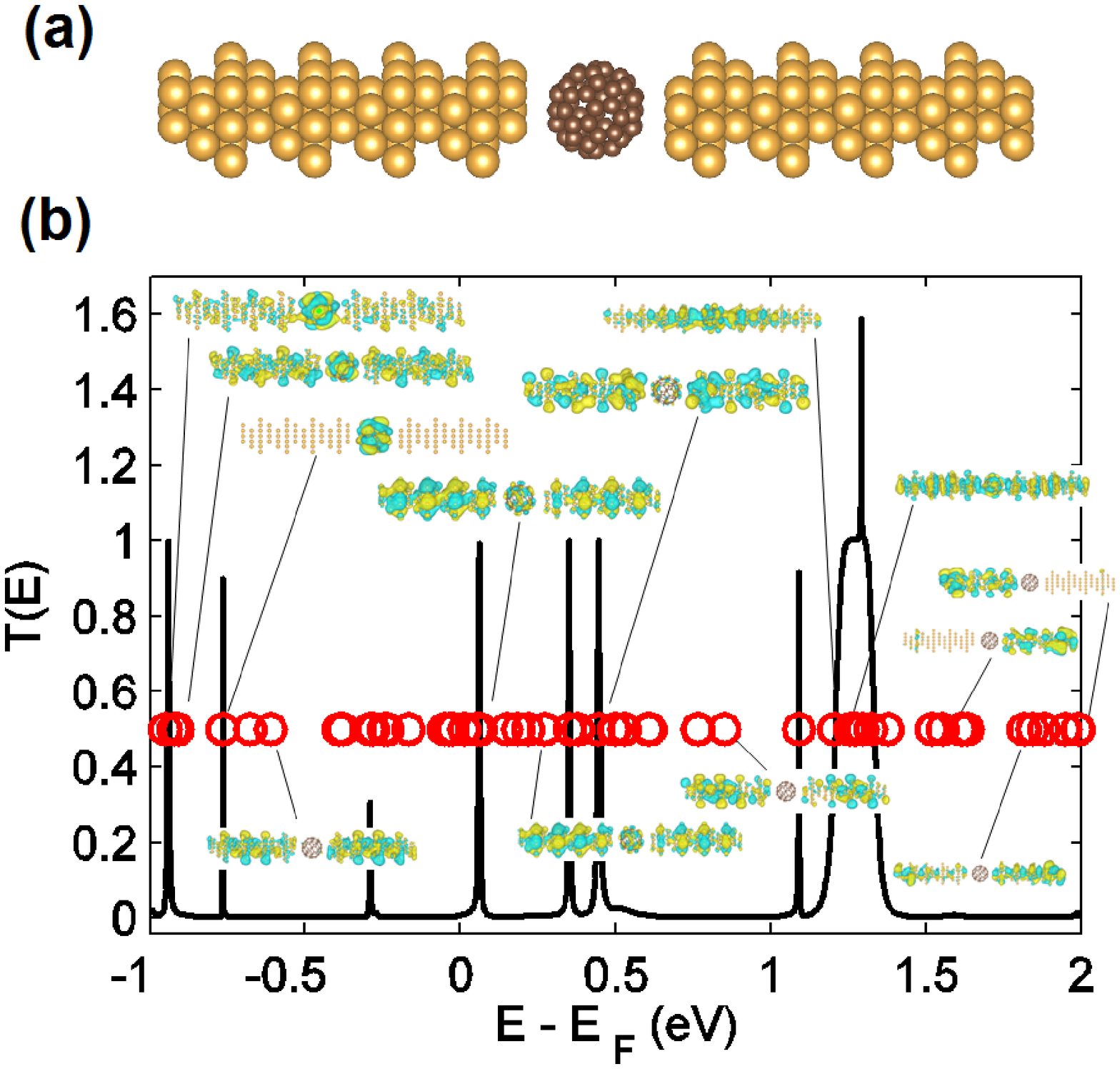} 
\caption{(a) The Au(111) nanowire/C$_{60}$/nanowire system.  The separation between the molecule and the Au(111) plane is 7.2 $a_0$.  
(b) TRANSEC calculated results for transmission, $T(E)$, for the Au(111)/C$_{60}$/Au(111) system, together with the locations of the (real) KS eigenvalues (red circles), and isosurfaces of representative KS orbitals.  
\label{fig:C60} }
\end{figure}

Having applied our method to a relatively simple molecular junction, we now demonstrate it on a more complex molecular scattering region.  
Fig.~\ref{fig:C60} shows the Au(111) leads of Sections~\ref{sec:Au-nanowire} and \ref{sec:BDT} together with a C$_{60}$ buckminsterfullerene molecule scattering region.  It also shows the computed $T(E)$ curve for this system.  

To validate these results, we again show a KS eigenvalue analysis, similar to those presented in the last two sections.  Here again, large $T(E)$ peaks coincide with delocalized molecular orbitals that support transport.  For example, the large multiple peak between $E_F + 1.2$ eV and $E_F + 1.5$ eV coincides with several orbitals with highly delocalized probability densities, two of which are shown in the figure.  
The peak near $E_F - 0.76$ eV may be associated with the corresponding eigenvalue at the same location. 
However, when examining its KS orbital, it seems to be highly localized on the C$_{60}$ molecule with little contribution from the Au lead sections.  Hence, we believe that this peak corresponds to the adjacent eigenvalue near $E_F - 1$ eV, which presents a delocalized orbital (also shown) that is more likely to support current.  

\subsection{Discussion and future work \label{sec:Discuss}}

Having demonstrated the reliability of our approach, we reiterate its advantages and weaknesses.  
We remark that 
the TRANSEC calculation can be considered as a benchmark to test other linear-response transport results using smaller basis sets, just as a converged real-space or plane-wave DFT calculation is customarily used to benchmark DFT results.  
In addition, the real-space method does not suffer from ghost-state transmission \cite{GhostTransRatner,ReuterPartition} and has other favorable convergence properties, as mentioned earlier.  

Methodologically, we note that the use of ABCs allows TRANSEC to simulate realistic electrodes with extensive contact regions at an affordable computational cost.  This is mainly due to avoiding the explicit calculation of the self-energies representing the semi-infinite leads, although the use of ABCs does require longer (and more realistic) contact regions.  Compared to the jellium approximation, the ABCs absorb outgoing waves, and are therefore more likely to avoid spurious reflections from the edges of the finite lead models.  In addition, as we noted in Sec.~\ref{sec:atom-chains}, 
tuning the ABC parameters to a set of electrodes is robust (although in general the parameters may depend on energy).  

We note that despite the real-space method's suggested status as a benchmark for other transport calculations, it remains a computationally expensive method when applied to small systems where significant parallelization is unnecessary.  
However, for large calculations, the number of complex eigenpairs needed (see discussion at the end of Section \ref{sec:cplx-diag}) can grow more slowly than the system size, so the problem becomes favorable.  The method's high parallelizability allows the fast evaluation (with sufficient processors) of transport problems that would be highly challenging with other representations.  As computational resources become more available with time, the real-space method should help meet the demand for more realistic transport calculations on large systems.  

One direction for further work is to implement a more efficient form \cite{RissMeyer-ABCs} for the ABCs than Eq.~(\ref{eq:ABC}).  As mentioned near Eq.~(\ref{eq:T-realspace}), only a sub-matrix of $G$ between the two ABCs is needed.  
Furthermore, the decay of the ABC itself influences the overall size of the simulation cell needed.  
Therefore, a form of ABC that occupies less volume in the simulation cell will considerably lower the computational demand.  

Another possible improvement that takes good advantage of the sparsity of $H$ is a divide and conquer method for transport.\cite{OdedDC, ReuterDC}  Previous work has shown that transport through a long, nearly-homogenous system can be computed in a time that scales linearly with the number of scatterers.  This approach has the added advantage that it works well with the natural parallelizability of the real-space method.

Performing self-consistent NEGF transport calculations (beyond linear response to bias) \cite{Datta, KeBaranger} is another direction for future work.  In our formalism this would pose several challenges.  As emphasized in the previous discussion 
(see also Sec.~\ref{sec:Method}), our algorithm takes advantage of 
the fact that evaluating $T$ 
requires only a subset of the possible Green's function elements.  This may not be the case when performing fully self-consistent NEGF.  
It typically requires the evaluation of a larger portion of $G$ as well as its integration over a very large energy range, in each self-consistent iteration.  
The latter integration would be a challenge because the larger energy domain could require finding more eigenpairs in Eq.~(\ref{eq:complex-energy-evp}).  Also, the wide-band approximation for the ABCs could break down, especially if complex energies are used for a contour integration.

\section{Conclusions \label{sec:Conclude}}

We have presented a real-space method, which we call TRANSEC, for highly parallelized, first-principles electronic transport calculations in nanostructures.  
We have also demonstrated the validity of the method with several applications, including both small and large systems.  These applications displayed good agreement with both reference calculations and analytical expectations.  Finally, we have considered some potential directions for future extensions and applications.

\section{Acknowledgments}

We thank Iliya Lichtzier (WIS) for valuable technical advice and support, and Yousef Saad (University of Minnesota), Ariel Biller (WIS), Ofer Sinai (WIS), and Amir Natan (TAU) for helpful discussions and advice.  Work at the Weizmann Institute was supported by the European Research Council, the Israel Science Foundation, and the Lise Meitner Center for Computational Chemistry.  
Work at TAU was supported by the Israel Science Foundation (ISF), the German-Israeli Fund under Research Grant No.2291-2259.5/2011, the European Community's Seventh Framework Programme FP7/2007–2013 under grant agreement No. 249225, the Center for Nanoscience and Nanotechnology at Tel-Aviv University, and the Lise Meitner-Minerva Center for Computational Quantum Chemistry.
T.S. is grateful to the US National Science Foundation (Grant No. CHE-1012207) and to the US Department of Energy (Grant No. DE-SC0001785) for support.

\appendix*

\section{Derivation of Eqs.~(\ref{eq:T-realspace}) and (\ref{eq:T-tilde-realsp})}

Starting from Eq.~(\ref{eq:T}) and using the diagonal form of $\Gamma_{L,R}$, we find 
\begin{eqnarray}
T\left( E \right) = \mbox{Tr}\left\{ G^r\left( E \right) \: \Gamma_R \: G^a\left( E \right) \: \Gamma_L \right\} \\ \nonumber
\equiv \; \sum_{i \: j \: k \: l}^N G^r_{i j} \; \Gamma_{R j k} \: G^a_{k l} \: \Gamma_{L l i}  \\ \nonumber
= \; \sum_{i j}^N G^r_{i j} \; \Gamma_{R j j} \: G^a_{j i} \: \Gamma_{L i i}  \; .
\end{eqnarray}
Applying $G^a = G^{r \dagger}$, we recover
\begin{eqnarray}
T\left( E \right) = \; \sum_{i j}^N G^r_{i j} \; \Gamma_{R j j} \: G^{r *}_{i j} \: \Gamma_{L i i}  \\ \nonumber 
= \; \sum_{i j}^N |G^r_{i j}|^2 \; \Gamma_{R j j} \: \Gamma_{L i i} \; .
\end{eqnarray}
Noting that, by definition, $\Gamma_{L i i} \neq 0$ only for $i \in L$, and similarly for $\Gamma_R$, which completes the derivation of Eq.~(\ref{eq:T-realspace}).  

The derivation of Eq.~(\ref{eq:T-tilde-realsp}) proceeds the same way, except that now $G$ is diagonal instead of $\Gamma$, and we substitute an explicit form for $G$.  Starting from Eq.~(\ref{eq:T-tilde}), we obtain 
\begin{eqnarray}
T\left( E \right) = \mbox{Tr}\left\{ \tilde{G}\left( E \right) \: \tilde{\Gamma}_R \: \tilde{G}^a\left( E \right) \: \tilde{\Gamma}_L \right\} \\ \nonumber
\equiv \; \sum_{i \: j \: k \: l}^N \tilde{G}^r_{k j} \; \tilde{\Gamma}_{R j i} \: \tilde{G}^a_{i l} \: \tilde{\Gamma}_{L l k}  \\ \nonumber
= \; \sum_{i j}^N \tilde{G}^r_{j j} \; \tilde{\Gamma}_{R j i} \: \tilde{G}^a_{i i} \: \tilde{\Gamma}_{L i j}  \; ,
\end{eqnarray}
where we used the diagonal property of $\tilde{G}$, $\tilde{G}^r_{k j} = \tilde{G}^r_{j j} \: \delta_{k j}$, to set $k = j$ and $l = i$.  
Finally, substituting Eq.~(\ref{eq:G-diag}) for $G^r$ and $G^a$, we obtain Eq.~(\ref{eq:T-tilde-realsp}).


\begin{thebibliography}{1}

\bibitem{Datta} S.~Datta, \textit{Electronic Transport in Mesoscopic Systems}, Cambridge University Press, 1995; S.~Datta, \textit{Nanotechnology}, \textbf{15} S433 (2004).  

\bibitem{Landauer} R.~Landauer, \textit{IBM J. Res. Dev.}, \textbf{32}, 306 (1988); \textit{Physica Scripta}, \textbf{T42}, 110 (1992). 

\bibitem{DiVentra} M.~DiVentra, \textit{Electrical Transport in Nanoscale Systems}, Cambridge University Press, 2008.  

\bibitem{Transiesta}  M.~Brandbyge, J.-L.~Mozos, P.~Ordej\'{o}n, J.~Taylor, and K.~Stokbro, \textit{Phys. Rev. B}, \textbf{65} 165401 (2002).  

\bibitem{Smeagol} I.~Rungger and S.~Sanvito, \textit{Phys. Rev. B}, \textbf{78} 035407 (2008); A.~R.~Rocha, V.~M.~Garc\'{i}a-su\'{a}rez, S.~W.~Bailey, C.~J.~Lambert, J.~Ferrer and S.~Sanvito, \textit{Nature Materials} \textbf{4}, 335 (2005).  

\bibitem{KeBaranger}
S.-H. Ke, H.~U.~Baranger, and W.~Yang, \textit{Phys. Rev. B} \textbf{70}, 085410 (2004).  

\bibitem{WanT}
A.~Calzolari, N.~Marzari, I.~Souza, and M.~B.~Nardelli,  \textit{Phys. Rev. B}, \textbf{69}, 035108 (2004).  

\bibitem{DiVentraBDT}
M.~Di Ventra, S.~T.~Pantelides, and N.~D.~Lang, \textit{Phys. Rev. Lett.}, \textbf{84}, 979 (2000).  

\bibitem{GhostTransRatner}
C. Herrmann, G. C. Solomon, J. E. Subotnik, V. Mujica, and M. A. Ratner, 
\textit{J. Chem. Phys.} \textbf{132}, 024103 (2010).

\bibitem{ReuterPartition}
M.~G.~Reuter and R.~J.~Harrison. \textit{J. Chem. Phys.}, \textbf{139}, 114104 (2013).

\bibitem{PlaneWaveTransport}
H.~J.~Choi and J.~Ihm, \textit{Phys. Rev. B} \textbf{59}, 2267 (1999); A.~Smogunov, A.~Dal Corso, E.~Tosatti, \textit{Phys. Rev. B} \textbf{70}, 045417 (2004).  

\bibitem{parsec-94}  J. R. Chelikowsky, N. Troullier, and Y. Saad, \textit{Phys. Rev. Lett.} \textbf{72}, 1240 (1994).

\bibitem{KronikParsec} L. Kronik, A. Makmal, M. L. Tiago, M. M. G. Alemany, M. Jain, X. Huang, Y. Saad, and J.R. Chelikowsky, \textit{Phys. Stat. Solidi (b)} \textbf{243}, 1063 (2006).

\bibitem{GFLEUR}
D.~Wortmann, H.~Ishida, and S.~Bl\"{u}gel, \textit{Phys. Rev. B}, \textbf{65} 165103 (2002);  \textit{Phys. Rev. B} \textbf{66}, 075113 (2002).  

\bibitem{ParsecBenchmarks} M.~M.~G.~Alemany, M.~Jain, M.~L.~Tiago, Y.~Zhou, Y.~Saad, and J.~R.~Chelikowsky, \textit{Computer Phys. Comm.} \textbf{177}, 339 (2007).  

\bibitem{ChelikowskyParsec} 
J.~R.~Chelikowsky, \textit{J. Phys. D: Appl. Phys.} \textbf{33} R33–R50 (2000).

\bibitem{NatanParsecPBC}
A.~Natan,  A.~Benjamini, D.~Naveh, L.~Kronik, M.~L.~Tiago, S.~P.~Beckman, and J.~R.~Chelikowsky, \textit{Phys. Rev. B} \textbf{78}, 075109 (2008).  

\bibitem{Hirose} K. Hirose, T. Ono, Y. Fujimoto, S. Tsukamoto, \textit{First-Principles Calculations in Real-Space Formalism}, Imperial College Press, 2005.  

\bibitem{FH} Y. Fujimoto and K. Hirose, \textit{Phys. Rev. B} \textbf{67}, 195315 (2003). 

\bibitem{OnoAtomWires} T. Ono and K. Hirose, \textit{Phys. Rev. B} \textbf{70}, 033403 (2004);  T. Sasaki, T. Ono, and K. Hirose \textit{Phys. Rev. E } \textbf{74}, 056704 (2006).  

\bibitem{OnoTransport} T. Ono, S. Tsukamoto, Y. Egami, and Y. Fujimoto, \textit{J. Phys.: Condens. Matter} \textbf{23}, 394203 (2011).   

\bibitem{Kong} L.~Kong, M.~L.~Tiago, and J.~R.~Chelikowsky, \textit{Phys. Rev. B} \textbf{73}, 1195118 (2006).  

\bibitem{Kong2007} L. Kong, J. R. Chelikowsky, J. B. Neaton, and S. G. Louie, \textit{Phys. Rev. B} \textbf{76}, 235422 (2007).  

\bibitem{Neuhauser-ABCs}
D.~Neuhauser and M.~Baer, \textit{J. Chem. Phys.} \textbf{90}, 4351 (1988).  

\bibitem{SeidemanMiller}
T.~Seideman and W.~H.~Miller, \textit{J. Chem. Phys.} \textbf{96}, 4412 (1992); T.~Seideman and W.~H.~Miller, \textit{J. Chem. Phys.} \textbf{97}, 2499 (1992).  

\bibitem{RissMeyer-ABCs}
U.~V.~Riss and H.-D.~Meyer, \textit{J. Chem. Phys.} \textbf{105}, 1409 (1996).  

\bibitem{Henderson-ABCs}
T.~M. Henderson, G.~Fagas, E.~Hyde, and J.~C.~Greer, \textit{J. Chem. Phys.} \textbf{125}, 244104 (2006).

\bibitem{Hod-ABCs}
O.~Hod, E.~Rabani, and R.~Baer, \textit{Acc. Chem. Res.} \textbf{39}, 109 (2006); O.~Hod, R.~Baer, and E.~Rabani, \textit{J. Am. Chem. Soc.} \textbf{127}, 1648 (2005); O.~Hod, R.~Baer, and E.~Rabani, \textit{J. Phys.: Condens. Matter} \textbf{20}, 383201 (2008).

\bibitem{ABC-SelfEn}
J.~A.~Driscoll and K.~Varga, \textit{Phys. Rev. B} \textbf{78}, 245118 (2008).  

\bibitem{Roi-ACDC}
R.~Baer, T.~Seideman, S.~Ilani, and D.~Neuhauser, \textit{J. Chem. Phys.} \textbf{120}, 3387 (2004).   

\bibitem{ABC-Transient}
L.~Zhang, J.~Chen, and J.~Wang, \textit{Phys. Rev. B} \textbf{87}, 205401 (2013).  

\bibitem{ABC-multiterm}
B.~G.~Cook, P.~Dignard, and K.~Varga, \textit{Phys. Rev. B} \textbf{83}, 205105 (2011).  

\bibitem{ABC-review}
K.~Varga, \textit{Phys. Status Solidi B} \textbf{246}, 1407 (2009). 

\bibitem{ParsecBook}
J.~R.~Chelikowsky, L.~Kronik, I.~Vasiliev, M.~Jain, and Y.~Saad, 
in C.~Le Bris, Ed., \textit{Handbook of Numerical Analysis - Volume X: Computational Chemistry} (Elsevier, Amsterdam), 613 (2003).

\bibitem{Pickett}
W.~E.~Pickett, \textit{Comp. Phys. Rep.} \textbf{9}, 115 (1989).  

\bibitem{Chelikowsky}
J.~R.~Chelikowsky, \textit{J. Phys. D: Appl. Phys.} \textbf{33} R33( 2000).  

\bibitem{SantraCederbaum}
R. Santra and L. S. Cederbaum, \textit{Phys. Rep.} \textbf{368}, 1 (2002).  

\bibitem{CSYM}
A.~Bunse-Gerstner and R.~St\"{o}ver, \textit{Lin. Alg. Appl.}, 287 (1999), 105.

\bibitem{Arpack}
http://www.caam.rice.edu/software/ARPACK/
http://forge.scilab.org/index.php/p/arpack-ng/

\bibitem{CSYM-scaling}
In principle, the complex-symmetric Lanczos \cite{SantraCederbaum, CSYM, Freund} iterative eigensolution algorithm should be more efficient than the Arnoldi algorithm implemented in ARPACK \cite{Arpack} 
because re-orthogonalization of the Lanczos basis is not performed in each iteration.  However, in practice, we find it to be less stable than ARPACK in duplicating or omitting eigenpairs.  This is an issue for future work.  

\bibitem{PRIMME}
A.~Stathopoulos and J.~R.~McCombs, \textit{ACM Trans. Math. Software}, \textbf{37}, 21 (2010).

\bibitem{Freund}
R. Freund, \textit{SIAM J. Sci. Stat. Comput.} \textbf{13}, 425 (1992).

\bibitem{CA}
D.~M.~Ceperly and B.~J.~Alder, \textit{Phys. Rev. Lett.} \textbf{45}, 566 (1980).

\bibitem{PBE}
J.~P.~Perdew, K.~Burke, and M.~Ernzerhof, \textit{Phys. Rev. Lett.} \textbf{77}, 3865 (1996).

\bibitem{TIMES}
D. Sharma, L. Ansari, B. Feldman, M. Iakovidis, J. Greer, and G. Fagas, \textit{J. Appl. Phys.} \textbf{113}, 203708 (2013).

\bibitem{OMX}
T.~Ozaki, \textit{Phys. Rev. B} \textbf{67}, 155108 (2003);  T.~Ozaki and H.~Kino, \textit{Phys. Rev. B} \textbf{69}, 195113 (2004).

\bibitem{TroullierMartins}
N.~Troullier and J.~L.~Martins, \textit{Phys. Rev. B} \textbf{43}, 1993 (1991).  

\bibitem{DoronLeeor}
D.~Naveh, L.~Kronik, M.~L.~Tiago and J.~R.~Chelikowsky, \textit{Phys. Rev. B} \textbf{76}, 153407 (2007).   

\bibitem{Stokbro-BDT}
K.~Stokbro, J.~Taylor, M.~Brandbyge, J.-L.~Mozos, and P. Ordej\'{o}n, \textit{Comp. Mat. Sci.} \textbf{27} 151 (2003).  

\bibitem{Emberly}
E.~G.~Emberly and G.~Kirczenow, \textit{Phys. Rev. B} \textbf{58}, 10911 (1998); E.~G.~Emberly and G.~Kirczenow, \textit{Nanotechnol.} \textbf{10} 285 (1999).

\bibitem{Li-BDT}
Z.~Li and D.~S.~Kosov, \textit{Phys. Rev. B} \textbf{76}, 035415 (2007).

\bibitem{Bauschlicher}
C.~W.~Bauschlicher Jr., J.~W.~Lawson, A.~Ricca, Y.~Xue, and M.~A.~Ratner, \textit{Chem. Phys. Lett.} \textbf{388} 427 (2004).  

\bibitem{Strange}
M.~Strange, I.~S.~Kristensen, K.~S.~Thygesen, and K.~W.~Jacobsen, \textit{J. Chem. Phys.} \textbf{128}, 114714 (2008). 


\bibitem{PartialInversion}
L. Lin, C. Yang, J. C. Meza, J. Lu, L. Ying, and W. E, \textit{ACM Trans. Math. Softw.} \textbf{37}, 4, Article 40 (2011).

\bibitem{OdedDC}
O.~Hod, J.~E.~Peralta, and G.~E.~Scuseria, \textit{J. Chem. Phys.} \textbf{125}, 114704 (2006).  

\bibitem{ReuterDC}
M.~G.~Reuter, T.~Seideman, and M.~A.~Ratner, \textit{Phys. Rev. B} \textbf{83}, 085412 (2011); M.~G.~Reuter and J.~C.~Hill, \textit{Comput. Sci. Disc.} \textbf{5} 014009 (2012).  

\end{thebibliography}
\end{document}